\documentstyle[12pt]{article} 
\newcommand{\sn}{{\rm sn}}
\newcommand{\cn}{{\rm cn}}
\newcommand{\dn}{{\rm dn}}

\begin{document} 

\vspace{30mm}  

\centerline{\Large \bf The Elliptic Function in Statistical Integrable Models II}

\vspace{15mm}

\centerline {{\it Tezukayama University , Tezukayama 7, Nara 631, Japan
}}

\centerline{Kazuyasu Shigemoto\footnote{E-mail address:
shigemot@tezukayama-u.ac.jp}}

\vspace{20mm}

\centerline{\bf Abstract} 

\vspace{3mm} 
Two dimensional statistical integrable models, such as the Ising model, 
the chiral Potts model and the Belavin model, becomes integrable.
Because of the $SU(2)$ symmetry of these models, these models become integrable.
The integral models are often parameterized by the elliptic function or the
elliptic theta function.
In this paper, we study the Ising model, and we show that 
the Yang-Baxter equation of the Ising model can be written 
as the integrability condition of $SU(2)$, and we gives the natural
explanation why the Ising model can be parameterized by the 
elliptic function by connecting the spherical trigonometry relation
and the elliptic function.
Addition formula of the elliptic function is the secret of the exact 
solvability of the Ising model.

\vspace{5mm}
\setcounter{equation}{0}
\section{Introduction}

There are many two dimensional integrable statistical models, which are 
classified into the spin model, the vertex model and the face model\cite{Baxter}.

First and famous integrable and exactly solvable model is the Ising model\cite{Onsager}.
The Ising model is the $2$-state model, and the generalized $N$-state integrable model 
is the chiral Potts model\cite{chiral}. The structure of the integrability 
condition of the 
Ising model is $SU(2)$. While that of the chiral Potts model is the cyclic 
$SU(2)$\cite{Shigemoto,Shigemoto1}.
The Boltzmann weight of the Ising model can be parameterized by the elliptic 
function.   

While the $2 \times 2$-state integrable vertex model is the $8$-vertex model\cite{Baxter1}, 
and the generalized $N \times N$-state integrable model is the Belavin model\cite{Belavin}.
In the Belavin model, the structure of the integrability condition is the cyclic 
$SU(2)$\cite{Tracy,Cherednik}.
The Boltzmann weight for the 8-vertex model is parameterized by the elliptic function, 
and that for the Belavin model by the elliptic theta function with characteristics.

Baxter's hard hexagon model\cite{Baxter2} is the first integrable $2$-state face model,
which is obtained from the $8$-vertex model by the vertex-face correspondence,
and the generalized integrable model is  $A^{(1)}_{N-1}$ model\cite{Jimbo}. The Boltzmann weight
for these models are parameterized by the elliptic theta function with characteristics.

Then the origin of the integrability condition of the two dimensional statistical model
comes from the $SU(2)$ symmetry or the variant of $SU(2)$ symmetry. 
And we have the elliptic representation of the Boltzmann weight in many important cases.
Then we expect the correspondence between the $SU(2)$ symmetry 
and the elliptic function. In other words, we expect that the symmetry 
of the elliptic function is the $SU(2)$ symmetry, and the origin of the 
addition formulae of the elliptic function comes from the $SU(2)$ group structure.
This is just the same as the origin of the trigonometric addition formulae 
comes from the $U(1)$ group structure.
For the trigonometric function, we make the connection with the circle, which 
has $U(1)$ symmetry.  For the elliptic function, we make the connection with 
the surface of the sphere, which has $SU(2)$ symmetry.

\vspace{10mm}

\setcounter{equation}{0}
\section{Integrability condition and exactly solvable condition}

Let's consider the general Yang-Baxter equation of the spin model, and we 
explain the difference between the integrability condition and the 
exactly solvable condition. The Yang-Baxter equation is given by  
\begin{eqnarray}
  && V(u_3)U(u_3,u_1) V(u_1)= U(u_1) V(u_1,u_3) U(u_3).
\label{e2-1}
\end{eqnarray}   
This type of Yang-Baxter equation is called the integrability condition.
The meaning of the integrability condition is that the model has the group 
structure so that the model has nice symmetry. As the Yang-Baxter equation 
of this type says that the product of three group action for two different path
gives the same group action, the group structure of 
the model is expected to be $SU(2)$ or the variation of $SU(2)$.

Furthermore if the Yang-Baxter equation satisfies the difference property 
such as 
\begin{eqnarray}
  && V(u_3)U(u_1+u_3) V(u_1)= U(u_1) V(u_1+u_3) U(u_3),
\label{e2-2}
\end{eqnarray}  
we call this type of equation as the exactly solvable condition. This is called 
the exactly solvable condition because the non-Abelian group action can be realized by 
the addition of the parameter $u_i$. Then the non-Abelian model is parameterized 
by the Abelian parameters $u_i$, which makes the model exactly solvable.
The non-linear addition formula of $u_i$ is the key relation that the 
non-Abelian integrable model becomes exactly solvable.

The Ising model is one of the examples of the exactly solvable model.
Then we pick up this Ising Model and explain the parameterization 
of the Ising model, which has the difference property, in a quite 
explicite way.
The star-triangle relation of the Ising model can be written 
in the form\cite{Shigemoto,Shigemoto1}
 \begin{eqnarray}
  &&\exp( L^*_3\sigma_x ) \exp( K_2\sigma_z ) \exp( L^*_1\sigma_x )
   =\exp( K_1\sigma_z ) \exp(L^*_2\sigma_x ) \exp( K_3\sigma_z ) .
\label{e2-3}
\end{eqnarray}

In the previous paper\cite{Shigemoto1}, we give the parameterization 
of the Ising model by the elliptic function in the form   
\begin{eqnarray}
  &&\cosh 2K_i=\frac{1}{\cn(u_i)} , \quad
  \sinh 2K_i=\frac{\sn(u_i)}{\cn(u_i)} ,
\label{e2-4}\\
  &&\cosh 2L^*_i=\frac{1}{\sn(K-u_i)} , \quad 
  \sinh 2L^*_i=\frac{\cn(K-u_i)}{\sn(K-u_i)}, 
\label{e2-5}\\
  &&(i=1,2,3)  .
\nonumber
\end{eqnarray}
Comparing Eq.(\ref{e2-1}) with Eq.(\ref{e2-3}), we have
\begin{eqnarray}
  &&U(u_i)=\exp( K_i\sigma_z ), \quad V(u_i)=\exp(L^*_i\sigma_x ) ,
\label{e2-6}
\end{eqnarray}  
and the Yang-Baxter equation comes to have the difference 
property, that is, we have the exactly solvable condition in the form 
\begin{eqnarray}
  && V(u_3)U(u_1+u_3) V(u_1)= U(u_1) V(u_1+u_3) U(u_3) ,
\label{e2-7}
\end{eqnarray} 
by the parameterization of the elliptic function.
We will explain why the Ising model can be parameterized by the 
elliptic function, and the Ising model comes to have the difference 
property.

\vspace{10mm}
\setcounter{equation}{0}
\section{Differential equation for the angle and the arc in the spherical triangle}

The connection between the addition formula of the elliptic function and the 
addition formula of the spherical triangle has the long history, which starts from
Lagrange and Legendre\cite{Lagrange,Legendre} and nice review article is given by 
Greenhill\cite{Greenhill}. 
We briefly review how to parameterize spherical angle and spherical arc 
by the elliptic function, which give the 
parameterization of the Yang-Baxter relation by the elliptic function.

\subsection{First differential equation}
In this subsection, we will give the differential equation for the angle on the 
sphere.
Various spherical trigonometry formulae are given by Todhunter\cite{Todhunter}.
The proof of the necessary
 formula we use here is given in the Appendix. \\
The first law of cosine is given by 
\begin{eqnarray}
  && \cos(a_1)=\cos(a_2) \cos(a_3)+\cos(A_1)\sin(a_2) \sin(a_3),
\label{e3-1}\\
  && \cos(a_2)=\cos(a_3) \cos(a_1)+\cos(A_2)\sin(a_3) \sin(a_1),
\label{e3-2}\\
  && \cos(a_3)=\cos(a_1) \cos(a_2)+\cos(A_3)\sin(a_1) \sin(a_2),
\label{e3-3}
\end{eqnarray}
and the law of sine is given by 
\begin{eqnarray}
  &&\frac{\sin(A_1)}{\sin(a_1)}
=\frac{\sin(A_2)}{\sin(a_2)}
=\frac{\sin(A_3)}{\sin(a_3)}=k .
\label{e3-4}
\end{eqnarray}
We keep $k$ to be constant, and we fix $a_3$, which means that we fix $A_3$ 
through Eq.(\ref{e3-4}).

Next we differentiate Eq.(\ref{e3-3}) and we have 
\begin{eqnarray}
  &&0=-\sin(a_1)\cos(a_2) da_1 -\cos(a_1) \sin(a_2) da_2 
\nonumber\\
 &&+\cos(A_3) \cos(a_1) \sin(a_2) da_1 +\cos(A_3) \sin(a_1) \cos(a_2) da_2
 \nonumber\\
&&= (-\sin(a_1)\cos(a_2)+\cos(A_3) \cos(a_1) \sin(a_2))da_1
\nonumber\\
  && +(-\cos(a_1) \sin(a_2) +\cos(A_3) \sin(a_1) \cos(a_2)) da_2  .
\label{e3-5}
\end{eqnarray}
Substituting expressions
\begin{eqnarray}
  &&\cos(A_1)=\frac{\cos(a_1)-\cos(a_2)\cos(a_3)}{\sin(a_2) \sin(a_3)} ,
\nonumber\\
 &&\cos(A_2)=\frac{\cos(a_2)-\cos(a_3)\cos(a_1)}{\sin(a_3) \sin(a_1)} ,
 \nonumber\\
&&\cos(A_3)=\frac{\cos(a_3)-\cos(a_1)\cos(a_2)}{\sin(a_1) \sin(a_2)} ,
\label{e3-6}
\end{eqnarray}
into the first term of Eq.(\ref{e3-5}), we have
\begin{eqnarray}
  &&({\rm First\ term})=(-\sin(a_1) \cos(a_2)+\cos(A_3) \cos(a_1) \sin(a_2)) da_1
\nonumber\\
 &&=\frac{-\sin^2(a_1) \cos(a_2)+\cos(a_1) \cos(a_3)-\cos^2(a_1) \cos(a_2)}
  {\sin{a_1}} da_1
\nonumber\\
  &&=\frac{-\cos(a_2)+\cos(a_1) \cos(a_3)}{\sin{a_1}} da_1
\nonumber\\
  &&=-\cos(A_2) \sin{a_3} da_1  .
\label{e3-7}
\end{eqnarray}
Similarly, for the second term of Eq.(\ref{e3-5}), we have
\begin{eqnarray}
  &&({\rm Second\ term})=(-\cos(a_1) \sin(a_2) +\cos(A_3) \sin(a_1) \cos(a_2)) da_2
\nonumber\\
 &&=-\cos(A_1) \sin{a_3} da_2  .
\label{e3-8}
\end{eqnarray}
Combining Eq.(\ref{e3-5}), Eq.(\ref{e3-7}), Eq.(\ref{e3-8}), we have
\begin{eqnarray}
  && \frac{da_1}{\cos(A_1)}+\frac{da_2}{\cos(A_2)}=0  .
\label{e3-9}
\end{eqnarray}
Using the relation $\sin(A_1)=k \sin(a_1)$, we have $\cos(A_1)=\sqrt{1-k^2 \sin^2(a_1)}$ 
where we assume $A_1$ is acute.
Then Eq.(\ref{e3-9}) can be written in the form
\begin{eqnarray}
  && \frac{da_1}{\sqrt{1-k^2 \sin^2(a_1)}}+\frac{da_2}{\sqrt{1-k^2 \sin^2(a_2)}}=0  .
\label{e3-10}
\end{eqnarray}
We put $x=\sin(a_1)$, $y=\sin(a_2)$, then we have $dx=\cos(a_1) da_1=\sqrt{1-x^2} da_1$,
 which gives $\displaystyle{da_1=\frac{dx}{\sqrt{1-x^2}}}$,
$\displaystyle{\frac{1}{\sqrt{1-k^2 \sin^2(a_1)}}=\frac{1}{\sqrt{1-k^2 x^2}}}$.
 Then Eq.(\ref{e3-10}) is written in the form of the differential equation of the
elliptic function 
\begin{eqnarray}
  && \frac{dx}{\sqrt{(1-x^2)(1-k^2 x^2)}}+\frac{dy}{\sqrt{(1-y^2)(1-k^2 y^2)}}=0  .
\label{e3-11}
\end{eqnarray}
Up to now, we consider $a_3$ to be fixed. For the general case, 
we put $z=\sin(a_3)$ and obtain the general differential equation of the form
\begin{eqnarray}
  && \frac{dx}{\sqrt{(1-x^2)(1-k^2 x^2)}}+\frac{dy}{\sqrt{(1-y^2)(1-k^2 y^2)}}
+\frac{dz}{\sqrt{(1-z^2)(1-k^2 z^2)}}=0  ,
\label{e3-12}
\end{eqnarray}
where we assume that  three angels $A_1$, $A_2$, $A_3$ are all acute.
If we put 
\begin{eqnarray}
  && u_1=\int^x \frac{dx}{\sqrt{(1-x^2)(1-k^2 x^2)}},
\label{e3-13}\\
  && u_2=\int^y \frac{dy}{\sqrt{(1-y^2)(1-k^2 y^2)}},
\label{e3-14}\\
  && u_3=\int^z \frac{dy}{\sqrt{(1-z^2)(1-k^2 z^2)}},
\label{e3-15}\\
  \nonumber\\
  && u_1+u_2+u_3={\rm Const.}, 
\label{e3-16}
\end{eqnarray}
we have $x=\sin(a_1)={\rm su}(u_1,k)$, that is, $a_1={\rm am}(u_1,k)$.
Similarly we have $a_2={\rm am}(u_2,k)$, $a_3={\rm am}(u_3,k)$.

\subsection{Second differential equation}
In this subsection, we will give the differential equation for the arc on the 
sphere.
We start from the second law of cosine 
\begin{eqnarray}
  && -\cos(A_1)=\cos(A_2) \cos(A_3)-\cos(a_1)\sin(A_2) \sin(A_3),
  \nonumber\\
  && -\cos(A_2)=\cos(A_3) \cos(A_1)-\cos(a_2)\sin(A_3) \sin(A_1),
  \nonumber\\
  && -\cos(A_3)=\cos(A_1) \cos(A_2)-\cos(a_3)\sin(A_1) \sin(A_2),
\label{e3-17}
\end{eqnarray}
and the law of sine
\begin{eqnarray}
  &&\frac{\sin(a_1)}{\sin(A_1)}
=\frac{\sin(a_2)}{\sin(A_2)}
=\frac{\sin(a_3)}{\sin(A_3)}=\frac{1}{k}.
\label{e3-18}
\end{eqnarray}

The differential equation for this system is obtained by replacing 
$a_i \rightarrow \pi-A_i$$(i=1,2,3)$ and $k\rightarrow 1/k$ in Eq.(\ref{e3-10}).
Then we put $X=\sin(A_1)$, $Y=\sin(A_2)$, $Z=\sin(A_3)$, and we obtain
\begin{eqnarray}
  && \frac{dX}{\sqrt{(1-X^2)(1-X^2/k^2)}}+\frac{dY}{\sqrt{(1-Y^2)(1-Y^2/k^2)}}
  \nonumber\\
 &&+\frac{dZ}{\sqrt{(1-Z^2)(1-Z^2/k^2)}}=0 ,
\label{e3-19}
\end{eqnarray}
where we assume that three length of arc $a_1$, $a_2$, $a_3$ are all acute.
If we put 
\begin{eqnarray}
  && U_1=\int^X \frac{dX}{\sqrt{(1-X^2)(1-X^2/k^2)}},
\label{e3-20}\\
  && U_2=\int^Y \frac{dY}{\sqrt{(1-Y^2)(1-Y^2/k^2)}},
\label{e3-21}\\
  && U_3=\int^Z \frac{dZ}{\sqrt{(1-Z^2)(1-Z^2/k^2)}},
\label{e3-22}\\
  \nonumber\\
  && U_1+U_2+U_3={\rm Const.},  
\label{e3-23} 
\end{eqnarray}
we have $A_i={\rm am}(U_i,1/k)$ $(i=1,2,3)$.
We also have another expression $X=\sin(A_i)={\rm sn}(U_i,1/k)=k{\rm sn}(U_i/k,k)$.
Using the law of sine, we have $k \sin(a_i)=\sin(A_i)$, and $\sin(a_i)={\rm sn}(u_i,k)$,
$\sin(A_i)=k {\rm sn}(U_i/k,k)$, we can connect $u_i$ and $U_i$ through 
${\rm sn}(u_i,k)={\rm sn}(U_i/k,k)$, so that we have $U_i=k u_i$, that is,
\begin{eqnarray}
  &&\sin(a_i)={\rm sn}(u_i,k) 
 \label{e3-24}\\
  &&\sin(A_i)={\rm sn}(k u_i,1/k)=k {\rm sn}(u_i,k) . 
 \label{e3-25}
\end{eqnarray}

We can obtain the addition formula of the elliptic function from the 
addition formula of the spherical cosine formula.
From Eq.(\ref{e3-2}) and Eq.(\ref{e3-17}), we have 
\begin{eqnarray}
  &&  \cos(a_2)=\cos(a_3) \cos(a_1)+\cos(A_2)\sin(a_3) \sin(a_1),
 \label{e3-26}\\
  && -\cos(A_2)=\cos(A_3) \cos(A_1)-\cos(a_2)\sin(A_3) \sin(A_1),
 \label{e3-27}
\end{eqnarray}
and elliminating $\cos(A_2)$ or $\cos(a_2)$, we have 
\begin{eqnarray}
  &&  \cos(a_2)
  =\frac{\cos(a_1) \cos(a_3)-\sin(a_1)\cos(A_1)\sin(a_3)\cos(A_3)}
      { 1-\sin(a_1)\sin(A_1)\sin(a_2)\sin(A_2)  },
 \label{e3-28}\\
  &&-\cos(A_2)
  =\frac{\cos(A_1) \cos(A_3)-\cos(a_1)\sin(A_1)\cos(a_3)\sin(A_3)}
      { 1-\sin(a_1)\sin(A_1)\sin(a_2)\sin(A_2)  },,
 \label{e3-29}
\end{eqnarray}

From here, we take $A_2$ to be obtude to find the addition formula 
of the elliptic function in the 
convenient form. In this case , we have $u_2=u_1+u_3$, and 
$\sin(a_i)={\rm sn}(u_i,k)$, $\sin(A_i)=k \sin(a_i)$, 
$\cos(a_i)=\cn(u_i,k)$ $(i=1,2,3)$, $\cos(A_1)=\dn(u_1,k)$, 
$\cos(A_2)=-\dn(u_2,k)$, $\cos(A_3)=\dn(u_3,k)$. 
Then Eq.(\ref{e3-28}) and Eq.(\ref{e3-29}) gives 
\begin{eqnarray}
  &&  \cn(u_2,k)=\cn(u_1+u_3,k) \nonumber\\
  && =\frac{\cn(u_1,k) \cn(u_3,k)-\sn(a_1,k)\dn(u_1,k)\sn(u_3,k)\dn(u_3,k)}
      { 1-k^2 \sn^2(u_1,k)\sn^2(u_3,k)} ,
 \label{e3-30}\\
  && \dn(u_2,k)=\dn(u_1+u_3,k) \nonumber\\
  &&=\frac{\dn(u_1,k) \dn(u_3,k)-k^2 \sn(u_1,k)\cn(u_1,k)\sn(u_3,k)\cn(u_3,k)}
      { 1-k^2 \sn^2(u_1,k)\sn^2(u_3,k)  } ,
 \label{e3-31}
\end{eqnarray}    
and  using $\sn(u_3,k)=\sqrt{1-\cn^2(u_3,k)}$, we have 
\begin{eqnarray}
  &&  \sn(u_2,k)=\sn(u_1+u_3,k) \nonumber\\
  && =\frac{\sn(u_1,k) \cn(u_3,k)\dn(u_3,k)+\cn(u_1,k)\dn(u_1,k)\sn(u_3,k)}
      { 1-k^2 \sn^2(u_1,k)\sn^2(u_3,k)}.
 \label{e3-32}
\end{eqnarray} 
These are convenient form of addition formula of the elliptic function. 

\vspace{10mm}
\setcounter{equation}{0}
\section{Parameterization by the elliptic function
} 
The spherical trigonometry relation such as laws of the first and 
the second cosine and the law of sines are obtained from the 
group theoretical relation of $SU(2)$ in the form 
\begin{eqnarray}
\exp\{iA_1 J_x\} \exp\{ia_2 J_z\} \exp\{iA_3 J_x\}
= \exp\{ia_3 J_z\} \exp\{i(\pi-A_2) J_x\} \exp\{ia_1 J_z\}  ,
\label{e4-1}
\end{eqnarray}
where $J_x$, $J_y$, $J_z$ are $SU(2)$ generators with 
$[J_x,J_y]=iJ_z$, $[J_y,J_z]=iJ_x$, $[J_z,J_x]=iJ_y$.
For $J=1$ case, Eq.(\ref{e4-1}) becomes
\begin{eqnarray}
&&\{1-\left(1-\cos(A_1)\right) J^2_x +i \sin(A_1) J_x \} 
\{1-\left(1-\cos(a_2)\right) J^2_z +i \sin(a_2) J_z \}
\nonumber\\
&&\{1-\left(1-\cos(A_3)\right) J^2_x +i \sin(A_3) J_x \} 
=\{1-\left(1-\cos(a_3)\right)J^2_z +i \sin(a_3) J_z \}
\nonumber\\
&&\{1-\left(1-\cos(\pi-A_2)\right)J^2_x +i \sin(\pi-A_2) J_x \}
\{1-\left(1-\cos(a_1)\right)J^2_z +i \sin(a_1) J_z \},
\nonumber\\
&& \nonumber\\
&&J_x=\left(\begin{array}{ccc}
           0 & 0& 0 \\
           0 & 0& -i \\
           0 & i& 0 
\end{array} \right) , \quad
J_z=\left(\begin{array}{ccc}
           0 & -i& 0 \\
           i & 0& 0 \\
           0 & 0& 0 
\end{array} \right) .
\nonumber
\end{eqnarray}\\
\unitlength 0.1in
\begin{picture}
(25,32)
(0,-40)
%
\special{pn 8}%
\special{ar 2980 2090 1419 1419  3.4952958 3.5071950}%
%
\special{pn 8}%
\special{ar 3310 2420 1850 930  3.5158448 4.4740680}%
%
\special{pn 8}%
\special{ar 800 2220 2190 2190  5.9570932 6.2831853}%
\special{ar 800 2220 2190 2190  0.0000000 0.5022463}%
%
\special{pn 8}%
\special{ar 3240 1980 1630 1380  1.8904448 3.0836330}%
%
\special{pn 8}%
\special{ar 1640 2080 263 263  5.5234226 6.2831853}%
\special{ar 1640 2080 263 263  0.0000000 1.4288993}%
%
\special{pn 8}%
\special{ar 2860 1520 298 298  1.2490458 2.9047439}%
%
\special{pn 8}%
\special{ar 2720 3290 295 295  3.5762629 5.1047291}%
\put(19.4000,-21.8000){\makebox(0,0)[lb]{A}}%
\put(20.4000,-22.3000){\makebox(0,0)[lb]{\scriptsize 3}}%
\put(25.4000,-20.0000){\makebox(0,0)[lb]{A}}%
\put(26.4000,-20.5000){\makebox(0,0)[lb]{\scriptsize 1}}%
\put(25.2000,-28.3000){\makebox(0,0)[lb]{A}}%
\put(26.2000,-28.8000){\makebox(0,0)[lb]{\scriptsize 2}}%
\put(17.5000,-30.4000){\makebox(0,0)[lb]{a}}%
\put(18.5000,-30.9000){\makebox(0,0)[lb]{\scriptsize 1}}%
\put(31.6000,-23.9000){\makebox(0,0)[lb]{a}}%
\put(32.6000,-24.4000){\makebox(0,0)[lb]{\scriptsize 3}}%

\put(19.9000,-15.8000){\makebox(0,0)[lb]{a}}%
\put(20.9000,-16.3000){\makebox(0,0)[lb]{\scriptsize 2}}%
%
\special{pn 8}%
\special{pa 1610 2070}%
\special{pa 1680 2600}%
\special{fp}%
\special{sh 1}%
\special{pa 1680 2600}%
\special{pa 1691 2531}%
\special{pa 1673 2547}%
\special{pa 1651 2537}%
\special{pa 1680 2600}%
\special{fp}%
%
\special{pn 8}%
\special{pa 2720 3290}%
\special{pa 3290 3460}%
\special{fp}%
\special{sh 1}%
\special{pa 3290 3460}%
\special{pa 3232 3422}%
\special{pa 3239 3445}%
\special{pa 3220 3460}%
\special{pa 3290 3460}%
\special{fp}%
%
\special{pn 8}%
\special{pa 2730 3280}%
\special{pa 3000 2800}%
\special{fp}%
\special{sh 1}%
\special{pa 3000 2800}%
\special{pa 2950 2848}%
\special{pa 2974 2846}%
\special{pa 2985 2868}%
\special{pa 3000 2800}%
\special{fp}%
%
\special{pn 8}%
\special{pa 2870 1540}%
\special{pa 2710 1060}%
\special{fp}%
\special{sh 1}%
\special{pa 2710 1060}%
\special{pa 2712 1130}%
\special{pa 2727 1111}%
\special{pa 2750 1117}%
\special{pa 2710 1060}%
\special{fp}%
%
\special{pn 8}%
\special{pa 2880 1520}%
\special{pa 3369 1487}%
\special{fp}%
\special{sh 1}%
\special{pa 3369 1487}%
\special{pa 3301 1472}%
\special{pa 3316 1491}%
\special{pa 3304 1511}%
\special{pa 3369 1487}%
\special{fp}%
%
\special{pn 8}%
\special{pa 1620 2060}%
\special{pa 1970 1680}%
\special{fp}%
\special{sh 1}%
\special{pa 1970 1680}%
\special{pa 1910 1715}%
\special{pa 1934 1719}%
\special{pa 1940 1743}%
\special{pa 1970 1680}%
\special{fp}%
\put(13.9000,-20.0000){\makebox(0,0)[lb]{P}}%
\put(14.9000,-20.5000){\makebox(0,0)[lb]{\scriptsize 3}}%
\put(30.3000,-14.6000){\makebox(0,0)[lb]{P}}%
\put(31.3000,-15.1000){\makebox(0,0)[lb]{\scriptsize 1}}%
\put(26.2000,-35.5000){\makebox(0,0)[lb]{P}}%
\put(27.2000,-36.0000){\makebox(0,0)[lb]{\scriptsize 2}}%
\put(14.1000,-24.4000){\makebox(0,0)[lb]{\bf n}}%
\put(15.1000,-24.9000){\makebox(0,0)[lb]{\bf \scriptsize 1}}%
\put(29.1000,-35.9000){\makebox(0,0)[lb]{\bf n}}%
\put(30.1000,-36.4000){\makebox(0,0)[lb]{\bf \scriptsize 2}}%
\put(29.4000,-31.8000){\makebox(0,0)[lb]{\bf n}}%
\put(30.4000,-32.3000){\makebox(0,0)[lb]{\bf \scriptsize 3}}%
\put(28.3000,-12.4000){\makebox(0,0)[lb]{\bf n}}%
\put(29.3000,-12.9000){\makebox(0,0)[lb]{\bf \scriptsize 4}}%
\put(16.3000,-18.1000){\makebox(0,0)[lb]{\bf n}}%
\put(17.3000,-18.6000){\makebox(0,0)[lb]{\bf \scriptsize 5}}%
\put(30.6000,-17.4000){\makebox(0,0)[lb]{\bf n}}%
\put(31.6000,-17.9000){\makebox(0,0)[lb]{\bf \scriptsize 6}}%
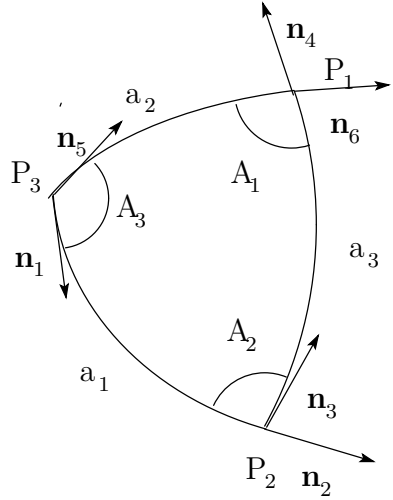
\put(10.1000,-38.0000){\makebox(0,0)[lb]{\bf Fig.1 The change of vectors on the spherical triangle}}%
\end{picture}%
\noindent \\
This relation is understand as the integrability condition 
on the sphere. We consider 3 points on the sphere $P_1$, $P_2$, $P_3$.
Starting from the vector ${\bf n_1}$ at point $P_3$. Then the vector parallel
transports along the tangential direction of the large circle or rotates 
around the point $P_i$$(i=1,2,3)$.
The following group element corresponds to the parallel transport or 
the rotation around the point $P_i$$(i=1,2,3)$ of the vector in the following way
\begin{eqnarray}
&&\exp\{iA_3 J_x\} : {\bf n_1} \rightarrow {\bf n_5}, \quad 
\exp\{ia_2 J_z\} : {\bf n_5} \rightarrow {\bf n_6}, \quad
\exp\{iA_1 J_x\} : {\bf n_6} \rightarrow {\bf n_4}, \nonumber\\
&&\exp\{ia_1 J_z\} : {\bf n_1} \rightarrow {\bf n_2}, \quad
\exp\{i(\pi-A_2) J_x\} : {\bf n_2} \rightarrow {\bf n_3}, \quad
\exp\{ia_3 J_z\}  :  {\bf n_3} \rightarrow {\bf n_4} \nonumber
\end{eqnarray}
Therefore Eq.(\ref{e4-1}) is the integrability condition that  
two different ways to change the vector from ${\bf n_1}$ to 
${\bf n_4}$, that is, path 1:$\{{\bf n_1} \rightarrow {\bf n_5} 
\rightarrow {\bf n_6}   \rightarrow {\bf n_4}\}$ 
and path 2:$\{{\bf n_1} \rightarrow {\bf n_2} \rightarrow {\bf n_3} 
  \rightarrow {\bf n_4}\}$,
gives the same group action, which is nothing but the integrability 
condition of $SU(2)$.

While the star-triangle relation can be written in the form 
\begin{eqnarray}
\exp\{2L^*_1 J_x\} \exp\{2K_2 J_z\} \exp\{2L^*_3 J_x\}
= \exp\{2K_3 J_z\} \exp\{2L^*_2 J_x\} \exp\{2K_1 J_z\}  .
\label{e4-2}
\end{eqnarray} 
In the previous paper, we parameterize the Ising model with the elliptic
function in the form 
\begin{eqnarray}
&& \cosh(2K_i)=\frac{1}{\cn(v_i,k')} , \quad \sinh(2K_i)=\frac{\sn(v_i,k')}{\cn(v_i,k')} ,
\label{e4-3}\\
&& \cosh(2L^*_i)=\frac{\dn(v_i,k')}{\cn(v_i,k')} , \quad 
\sinh(2L^*_i)= \frac{k \sn(v_i,k')}{\cn(v_i,k')} ,
\label{e4-4}\\
&& v_2=v_1+v_3, \quad (i=1,2,3)  .
\label{e4-5}
\end{eqnarray} 
By using the Jacobi's imaginary transformation, 
we put $v_i=i u_i$ and we have
\begin{eqnarray}
&& \sn(v_i,k')=i\frac{ \sn(u_i,k)}{ \cn(u_i,k)}, \quad
\cn(v_i,k')=\frac{1}{ \cn(u_i,k)}, \quad
\dn(v_i,k')=\frac{\dn(u_i,k)}{ \cn(u_i,k)}, \quad
\label{e4-6}
\end{eqnarray}
then we have 
\begin{eqnarray}
&& \cosh(2K_i)=\cn(u_i,k) , \quad \sinh(2K_i)=i \sn(u_i,k) ,
\label{e4-7}\\
&& \cosh(2L^*_i)=\dn(u_i,k) , \quad 
\sinh(2L^*_i)=i k \sn(u_i,k) ,
\label{e4-8}\\
&& u_2=u_1+u_3, \quad (i=1,2,3)  .
\label{e4-9}
\end{eqnarray} 
Using the relation
\begin{eqnarray}
&& \sn(u,1/k)=k \sn(u/k,k) ,\  
   \cn(u,1/k)=\dn(u/k,k) ,\nonumber\\  
 &&  \dn(u,1/k)=\cn(u/k,k) ,
\label{e4-10}
\end{eqnarray}
and putting $K_i=i\hat{K}_i$, $L^*_i=i\hat{L}^*_i$, we have
\begin{eqnarray}
&& \cos(2\hat{K}_i)=\cn(u_i,k) , \quad \sin(2\hat{K}_i)=\sn(u_i,k) ,
\label{e4-11}\\
&& \cos(2\hat{L}^*_i)=\cn(k u_i,1/k) , \quad 
\sin(2\hat{L}^*_i)=\sn(k u_i,1/k) ,
\label{e4-12}\\
&& u_2=u_1+u_3, \quad (i=1,2,3)   .
\label{e4-13}
\end{eqnarray} 
For the spherical triangle, we take $A_2$ as obtuse, and $A_1$ and $A_3$ are 
acute. In this case 
\begin{eqnarray}
  && \frac{dA_2}{\sqrt{1-\sin^2(A_2)/k^2}}
  =\frac{dA_1}{\sqrt{1- \sin^2(A_1)/k^2}}+\frac{dA_3}{\sqrt{1-\sin^2(A_3)/k^2}} ,
\label{e4-14}
\end{eqnarray}
and we take $U_2=U_1+U_3$, which gives $u_2=u_1+u_3$ by using $U_i=ku_i$$(i=1,2,3)$.
While the spherical triangle is parameterized by 
\begin{eqnarray}
&& \cos(a_i)=\cn(u_i,k) , \quad \sin(a_i)=\sn(u_i,k) ,
\label{e4-15}\\
&& \cos(A_i)=\cn(U_i,1/k)=\cn(k u_i,1/k), \quad 
\sin(A_i)=\sn(k u_i,1/k) ,
\label{e4-16}\\
&& u_2=u_1+u_3, \quad (i=1,2,3)  .
\label{e4-17}
\end{eqnarray}
Then we have 
\begin{eqnarray}
&& 2\hat{K}_i=a_i={\rm am}(u_i,k) , \quad
   2\hat{L}^*_i=A_i={\rm am}(k u_i,1/k)  , 
\label{e4-18}
\end{eqnarray}
where we must notice that $A_2$ is obtuse, so that $\pi-A_2$ is acute. 
Then the Yang-Baxter equation is written in the form
\begin{eqnarray}
&&\exp\{i {\rm am}(k u_1,1/k) J_x\} \exp\{i{\rm am}(u_2,k) J_z\} 
\exp\{i{\rm am}(k u_3,1/k) J_x\}
\nonumber\\
 &&= \exp\{i{\rm am}(u_3,k) J_z\} \exp\{i{\rm am}(k u_2,1/k) J_x\} 
\exp\{i{\rm am}(u_1,k) J_z\}  .
\label{e4-19}
\end{eqnarray}

\vspace{10mm}
\section{Abel's addition theorem} 
\setcounter{equation}{0}

The addition theorem of the algebraic function, that is, the Abel's 
addition theorem is the essential key relation to solve the statistical
model.
Then we consider the old problem of the Abel's addition theorem here.
We put $f_4(x)=A_0+A_1x+A_2x^2+A_3 x^3+ A_4 x^4$, and we consider 
the $g=1$ algebraic curve $y^2=f_4(x)$, 
then Jacobi or Richelot\cite{Jacobi,Richelot} tried to find the conserved quantity 
of the differential equation
\begin{eqnarray}
  &&\frac{dx_1}{\sqrt{f_4(x_1)}}+ \frac{dx_2}{\sqrt{f_4(x_2)}}=0  .
\label{e5-1} 
\end{eqnarray}
While we consider here the Abel's addition theorem of the standard type, that is,
\begin{eqnarray}
  &&\frac{dx_1}{\sqrt{f_4(x_1)}}+ \frac{dx_2}{\sqrt{f_4(x_2)}}
+\frac{dx_3}{\sqrt{f_4(x_3)}}=0  .
\label{e5-2} 
\end{eqnarray}
In general, we take 
\begin{eqnarray}
f_{2n-2}(x)=A_0+A_1x+A_2x^2+\cdots +A_{2n-3} x^{2n-3}+ A_{2n-2} x^{2n-2}  ,
\label{e5-3} 
\end{eqnarray}
and consider the differential equation of the form 
\begin{eqnarray}
  &&\frac{dx_1}{\sqrt{f_{2n-2}(x_1)}}+ \frac{dx_2}{\sqrt{f_{2n-2}(x_2)}}+ \cdots 
   +\frac{dx_n}{\sqrt{f_{2n-2}(x_n)}}=0 ,
\nonumber\\
  &&\frac{x_1 dx_1}{\sqrt{f_{2n-2}(x_1)}}+ \frac{x_2 dx_2}{\sqrt{f_{2n-2}(x_2)}}+ \cdots 
  + \frac{x_n dx_n}{\sqrt{f_{2n-2}(x_n)}}=0 , 
\nonumber\\
   &&  \cdots 
\nonumber\\
 &&\frac{{x_1}^{n-3} dx_1}{\sqrt{f_{2n-2}(x_1)}}
 + \frac{{x_2}^{n-3} dx_2}{\sqrt{f_{2n-2}(x_2)}}+ \cdots 
  + \frac{{x_n}^{n-3} dx_n}{\sqrt{f_{2n-2}(x_n)}}=0 .
\label{e5-4}
\end{eqnarray} 
We define $F(x)$ by  
\begin{eqnarray}
  &&F(x)=(x-x_1)(x-x_2) \cdots (x-x_{n-1})(x-x_n) , 
\label{e5-5}
\end{eqnarray}
then we have  
\begin{eqnarray}
  &&F'(x_1)=(x_1-x_2)(x_1-x_3) \cdots (x_1-x_{n-1})(x_1-x_n) .
\label{e5-6}
\end{eqnarray}
etc. 
Then if $x_i$, which depend on $t$, satisfies the following equation
\begin{eqnarray}
  &&\frac{dx_i}{dt}=\frac{x_i \sqrt{f_{2n-2}(x_i)}}{F'(x_i)}  ,
\label{e5-7}
\end{eqnarray}
we can show that Eq.(\ref{e5-4}) is satisfied. Here we use the following theorem for $F(x)$
\begin{eqnarray}
  &&\frac{x^k}{F(x)}=\sum_{i=1}^n \frac{x^k_i}{F'(x_i)} \frac{1}{(x-x_i)},\quad
(k=0,1,\cdots, n-1)  .  
\label{e5-8}
\end{eqnarray}
The proof is the followings:
As $F(x)$ is the $n$-th polynomial and it is much higher polynomial 
than $x^k$, we can write   
\begin{eqnarray}
  &&\frac{x^k}{F(x)}=\sum_{i=1}^n \frac{a_i}{(x-x_i)},\quad (k=0,1,\cdots, n-1)
\label{e5-9}  , 
\end{eqnarray}
then we multiply $(x-x_j)$ and take the limit $x \rightarrow x_j$, we have
\begin{eqnarray}
  &&\frac{x^k_j}{F'(x_j)}=\sum_{i=1}^n a_i \delta_{i,j}=a_j  ,
\label{e5-10}
\end{eqnarray} 
which gives Eq.(\ref{e5-8}).
By multiplying $x$ and taking $x \rightarrow \infty$ in Eq.(\ref{e5-8}), we have 
\begin{eqnarray}
  &&\sum_{i=1}^n \frac{x^k_i}{F'(x_i)} =\delta_{k,n-1},\quad
(k=0,1,\cdots, n-1)  . 
\label{e5-11}
\end{eqnarray}
Then the differential equation of the problem is satisfied in the following way
\begin{eqnarray}
 &&\sum^n_{i=1} \frac{{x_i}^{k} dx_i}{\sqrt{f_{2n-2}(x_i)}}
  =\sum^n_{i=1} \frac{{x_i}^{k+1} \sqrt{f_{2n-2}(x_i)} dt}{F'(x_i)\sqrt{f_{2n-2}(x_i)}}
  =\sum^n_{i=1} \frac{{x_i}^{k+1} dt}{F'(x_i)} =0, 
\label{e5-12}\\
&&(k=0,1,\cdots, n-3)
\nonumber
\end{eqnarray} 
\subsection{Abel's addition theorem I: one conserved quantity}
We will integrate this differential equation and find the conserved quantity.
Starting from Eq.(\ref{e5-4}), we have 
\begin{eqnarray}
  \frac{d^2 x_1}{dt^2}
  &=&\frac{dx_1}{dt} \frac{d}{dx_1}\left(\frac{x_1 \sqrt{f_{2n-2}(x_1)}}{F'(x_1)}\right)
 +\frac{dx_2}{dt} \frac{d}{dx_2}\left(\frac{x_1 \sqrt{f_{2n-2}(x_1)}}{F'(x_1)}\right)
  +\cdots 
\nonumber\\ 
 && +\frac{dx_n}{dt} \frac{d}{dx_n}\left(\frac{x_1 \sqrt{f_{2n-2}(x_1)}}{F'(x_1)}\right)
\label{e5-13}\\
&&=\frac{x_1 \sqrt{f_{2n-2}(x_1)}}{F'(x_1)}
 \frac{d}{dx_1}\left(\frac{x_1 \sqrt{f_{2n-2}(x_1)}}{F'(x_1)}\right)
+\frac{x_1 x_2 \sqrt{f_{2n-2}(x_1) f_{2n-2}(x_2)}}{F'(x_1) F'(x_2)} \frac{1}{(x_1-x_2)}
+\cdots
\nonumber\\
&& +\frac{x_1 x_n \sqrt{f_{2n-2}(x_1) f_{2n-2}(x_n)}}{F'(x_1) F'(x_n)} \frac{1}{(x_1-x_n)}
\label{e5-14}\\
&&=\frac{1}{2}\frac{d}{dx_1}\left(\frac{x_1^2 f_{2n-2}(x_1)}{(F'(x_1))^2}\right)
+\sum_{i=1}^n \frac{x_1 x_i \sqrt{f_{2n-2}(x_1) f_{2n-2}(x_i)}}{F'(x_1) F'(x_i)} \frac{1}{(x_1-x_i)}  .
\label{e5-15}
\end{eqnarray}
By taking the sum, we have
\begin{eqnarray}
 &&\frac{d^2 }{dt^2} \sum_{i=1}^n x_i=
 \frac{1}{2} \sum_{i=1}^n 
  \frac{d}{dx_i}\left(\frac{x_i^2 f_{2n-2}(x_i)}{F'(x_i)^2}\right)  .
\label{e5-16}
\end{eqnarray}
While we use the formula
\begin{eqnarray}
 &&\frac{x^2 f_{2n-2}(x) }{F(x)^2} -A_{2n-2}
  =\sum_{i=1}^n \left( \frac{x^2_i f_{2n-2}(x_i) }{F'(x_i)^2 (x-x_i)^2}
        +\frac{d}{dx_i}\left(  \frac{x^2_i f_{2n-2}(x_i) }{F'(x_i)^2} \right) 
        \frac{1}{(x-x_i)} \right)  .
\label{e5-17} 
\end{eqnarray}
The proof of the above is the following:
We can write
\begin{eqnarray}
 &&\frac{x^2 f_{2n-2}(x) }{F(x)^2} -A_{2n-2}
  =\sum_{i=1}^n \left( \frac{a_i}{(x-x_i)^2}+\frac{b_i}{(x-x_i)} \right)  .
\label{e5-18} 
\end{eqnarray}
We multiply $(x-x_j)^2$ and take the limit $x \rightarrow x_j$, the we have 
\begin{eqnarray}
 &&\frac{x^2_j f_{2n-2}(x_j) }{F'(x_j)^2}
  =\sum_{i=1}^n a_i \delta_{i,j} =a_j  .
\label{e5-19} 
\end{eqnarray}
Next we multiply $(x-x_j)^2$ and further multiply $\displaystyle{\frac{d}{dx}}$ and take the 
limit $x \rightarrow x_j$, we have 
\begin{eqnarray}
 && \lim_{x \rightarrow x_j} \frac{d}{dx}\frac{(x-x_j)^2 x^2 f_{2n-2}(x) }{F(x)^2}
  =\frac{d}{dx_j}\frac{x^2_j f_{2n-2}(x_j) }{F'(x_j)^2}=\sum_{i=1}^n b_i \delta_{i,j} =b_j .
\label{e5-20} 
\end{eqnarray}
From Eq.(\ref{e5-18}),  Eq.(\ref{e5-19}),  Eq.(\ref{e5-20}), we have 
 Eq.(\ref{e5-17}).
While by using the relation 
\begin{eqnarray}
 &&\frac{x^2 f_{2n-2}(x) }{F(x)^2} -A_{2n-2}
  =\left(A_{2n-3} +2A_{2n-2}\sum_{i=1}^n x_i \right) \frac{1}{x} +O\left(\frac{1}{x^2}\right)  ,
\label{e5-21} 
\end{eqnarray}
and by comparing the coefficient of $\displaystyle{O(\frac{1}{x})}$ in Eq.(\ref{e5-17}),
we have 
\begin{eqnarray}
 && A_{2n-3} +2A_{2n-2} \sum_{i=1}^n x_i= 
 \sum_{i=1}^n \frac{d}{dx_i}\left(  \frac{x^2_i f_{2n-2}(x_i) }{F'(x_i)^2} \right) 
  =2 \frac{d^2 }{dt^2} \sum_{i=1}^n x_i , 
\label{e5-22} 
\end{eqnarray}
where we use Eq.(\ref{e5-16}).
We put $\displaystyle{p=\sum_{i=1}^n x_i}$, then we have the differential equation of the form
\begin{eqnarray}
 && 2 \frac{d^2 p }{dt^2}=A_{2n-3} +2A_{2n-2} p  .
\label{e5-23} 
\end{eqnarray}
Multiplying $\displaystyle{\frac{dp}{dt}}$ in both side of Eq.(\ref{e5-23}), we have 
\begin{eqnarray}
 &&  \frac{d }{dt} \left(\left(\frac{dp}{dt}\right)^2- A_{2n-3} p -A_{2n-2} p^2 \right)=0  .
\label{e5-24} 
\end{eqnarray}
 Then we have one conserved quantity
\begin{eqnarray}
 &&  \left(\sum^n_{i=1} \frac{x_i \sqrt{f_{2n-2}(x_i)}}{F'(x_i)}\right)^2
=A_{2n-3} \sum_{i=1}^n x_i +A_{2n-2} \left(\sum_{i=1}^n x_i\right)^2 +C_1  .
\label{e5-25} 
\end{eqnarray}
\subsection{Abel's addition theorem II: another conserved quantity}
As Richelot\cite{Richelot} showed, we transform the original differential equation 
of $ x_i$ into that of  $\displaystyle{\xi_i=\frac{1}{x_i}}$. 
In this variable, we have 
\begin{eqnarray}
 &&  dx_i=-\frac{d\xi_i}{\xi^2_i} ,
\label{e5-26} \\
 && \frac{1}{\sqrt{f_{2n-2}(x_i)}}=\frac{1}{\sqrt{f_{2n-2}(1/\xi_i)}}
    =\frac{ \xi_i^{n-1}}{\sqrt{ g_{2n-2}(\xi_i)}} , 
\label{e5-27} \\
&&{\rm where}
\nonumber\\
&& g_{2n-2}(\xi_i)=A_{2n-2}+A_{2n-3}\xi+\cdots+A_{1}\xi^{2n-3}+A_{0}\xi^{2n-2}  , 
\label{e5-28}
\end{eqnarray}
which gives
\begin{eqnarray}
&& \frac{x_{i}^k dx_i}{\sqrt{f_{2n-2}(x_i)}}\ (k=0,1, \cdots, n-3)
 \nonumber\\
 &&=- \frac{\xi_{i}^{k'} d \xi_i} {\sqrt{g_{2n-2}(\xi_i)}} (k'=n-3-k=0,1, \cdots, n-3) .
\label{e5-29}
\end{eqnarray}
Therefore the differential equation is transformed into the form
\begin{eqnarray}
 &&\sum_{i=1}^n \frac{\xi_{i}^{k} d \xi_i} {\sqrt{g_{2n-2}(\xi_i)}}=0,\quad
  (k=0,1, \cdots, n-3)  . 
\label{e5-30}
\end{eqnarray}
Defining $F_1(\xi)=\prod_{i=1}^n (\xi-\xi_i)$ and comparing with Eq.(\ref{e5-25}), 
we have another conserved quantity 
\begin{eqnarray}
 &&  \left(\sum^n_{i=1} \frac{\xi_i \sqrt{g_{2n-2}(x_i)}}{F'_1(\xi_i)}\right)^2
=A_{1} \sum_{i=1}^n \xi_i +A_{0} (\sum_{i=1}^n \xi_i)^2 +C_2  .
\label{e5-31} 
\end{eqnarray} 
Using 
\begin{eqnarray}
 && F'_1(\xi_i)=\frac{(-1)^{n-1}}{x_1 x_2 \cdots x_n} 
  \frac{F'(x_i)}{x^{n-2}_i }  ,
\label{e5-32} 
\end{eqnarray} 
we have 
\begin{eqnarray}
 &&  \left(\sum^n_{i=1} \frac{\sqrt{f_{2n-2}(x_i)}}{x^2_iF'(x_i)}\right)^2 (x_1 x_2 \cdots x_n)^2
=A_{1} \sum_{i=1}^n \frac{1}{x_i} +A_{0}  (\sum_{i=1}^n \frac{1}{x_i})^2+C_2  .
\label{e5-33} 
\end{eqnarray}
\subsection{Simple $n=3$ case: Abel's addition theorem for the elliptic function}
For $n=3$ case with
$f_4(x)=(1-x^2)(1-k^2 x^2)$, we  numerically checked by Maxima the following 
addition formula 
\begin{eqnarray}
 &&  \sum^3_{i=1} \frac{x_i \sqrt{f_{4}(x_i)}}{F'(x_i)}
     +k\sum_{i=1}^3 x_i=0, 
 \label{e5-34}\\ 
&& (x_1 x_2 x_3) \sum^3_{i=1} \frac{\sqrt{f_{4}(x_i)}}{x^2_iF'(x_i)}
 -\sum_{i=1}^3 \frac{1}{x_i}=0,
\label{e5-35}\\ 
&&{\rm where}\nonumber\\
&&x_i=\sn(u_i), \quad \sqrt{f_{4}(x_i)})=\cn(u_i)\dn(u_i), \quad u_1+u_2+u_3=0 .
\nonumber
\end{eqnarray}

\section{Summary and discussion}

The integrability condition, which is called the Yang-Baxter equation,
 in two dimensional statistical models means that such integrable models 
have the group structure. The Boltzmann 
weight of such integrable models often can be parameterized by the 
elliptic function or the elliptic theta function.
Furthermore if the Yang-Baxter equation have the nice difference 
property, we can exactly solve the model. Such exactly solvable models
often can be parameterized by the elliptic function.
If we notice that the Yang-Baxter relation is the relation 
of the products of the three group action, we expect that 
the integrable model has the $SU(2)$ or $SU(2)$ variant group structure. 
Furthermore if the model is exactly solvable, we expect that the exactly 
solvable model can be parameterized by the elliptic function.
As the Ising model is the exactly solvable model, the above 
expectation is realized by rewriting 
the Yang-Baxter relation in the form of the integrability condition 
of $SU(2)$ group. Furthermore we can parameterize the Boltzmann weight
by the elliptic function, which comes to have the difference 
property, by making the connection between the $SU(2)$ group element 
and the elliptic function.

In this way, we can understand the reason why the Ising model
can be parameterized by the elliptic function in a quite 
natural way, where the connection between the $SU(2)$ 
and the elliptic function is essential.
Addition formula of the elliptic function is the secret of the exact 
solvability of the Ising model.

Finally, we find some of the conserved quantity for the 
hyperelliptic differential equation, which gives the explicit form of 
the Abel's addition theorem.
This will will be useful to
find the group structure of such hyperelliptic function.

\vspace{15mm}



\newpage

\appendix
\setcounter{equation}{0}
\section{\large \bf Spherical trigonometry formulae}

In this appendix, we give the brief proof of various 
spherical trigonometry relations.
We consider the sphere with unit radius and we put three
points $P_1$, $P_2$, $P_3$ on the sphere. We consider three unit vectors
which go from the origin to points $P_1$, $P_2$, $P_3$, 
and we denote such unit vecotors as ${\bf n_1}$, ${\bf n_2}$, ${\bf n_3}$ .
Further, we denote the length of the arc 
connecting ${\bf n_2}$ and ${\bf n_3}$
as $a_1$, and that connecting ${\bf n_3}$ and ${\bf n_1}$
as $a_2$, and that connecting ${\bf n_1}$ and ${\bf n_2}$ 
as $a_3$ . 

\unitlength 0.1in
\begin{picture}
(25,32)
(0,-40)
%
\special{pn 8}%
\special{ar 2980 2090 1419 1419  3.4952958 3.5071950}%
%
\special{pn 8}%
\special{ar 3310 2420 1850 930  3.5158448 4.4740680}%
%
\special{pn 8}%
\special{ar 800 2220 2190 2190  5.9570932 6.2831853}%
\special{ar 800 2220 2190 2190  0.0000000 0.5022463}%
%
\special{pn 8}%
\special{ar 3240 1980 1630 1380  1.8904448 3.0836330}%
%
\special{pn 8}%
\special{ar 1640 2080 263 263  5.5234226 6.2831853}%
\special{ar 1640 2080 263 263  0.0000000 1.4288993}%
%
\special{pn 8}%
\special{ar 2860 1520 298 298  1.2490458 2.9047439}%
%
\special{pn 8}%
\special{ar 2720 3290 295 295  3.5762629 5.1047291}%
\put(19.4000,-21.8000){\makebox(0,0)[lb]{A}}%
\put(20.4000,-22.3000){\makebox(0,0)[lb]{\scriptsize 3}}%
\put(25.4000,-20.0000){\makebox(0,0)[lb]{A}}%
\put(26.4000,-20.5000){\makebox(0,0)[lb]{\scriptsize 1}}%
\put(25.2000,-28.3000){\makebox(0,0)[lb]{A}}%
\put(26.2000,-28.8000){\makebox(0,0)[lb]{\scriptsize 2}}%
\put(17.5000,-30.4000){\makebox(0,0)[lb]{a}}%
\put(18.5000,-30.9000){\makebox(0,0)[lb]{\scriptsize 1}}%
\put(31.6000,-23.9000){\makebox(0,0)[lb]{a}}%
\put(32.6000,-24.4000){\makebox(0,0)[lb]{\scriptsize 3}}%

\put(19.9000,-15.8000){\makebox(0,0)[lb]{a}}%
\put(20.9000,-16.3000){\makebox(0,0)[lb]{\scriptsize 2}}%

\put(13.9000,-20.0000){\makebox(0,0)[lb]{P}}%
\put(14.9000,-20.5000){\makebox(0,0)[lb]{\scriptsize 3}}%
\put(30.3000,-14.6000){\makebox(0,0)[lb]{P}}%
\put(31.3000,-15.1000){\makebox(0,0)[lb]{\scriptsize 1}}%
\put(26.2000,-35.5000){\makebox(0,0)[lb]{P}}%
\put(27.2000,-36.0000){\makebox(0,0)[lb]{\scriptsize 2}}%
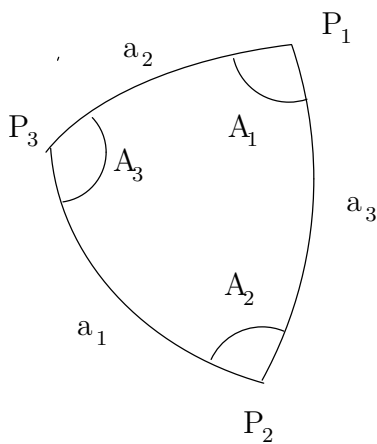
\put(10.1000,-38.0000){\makebox(0,0)[lb]{\bf Fig.2 Angles and arcs of the spherical triangle}}%
\end{picture}%
\noindent \\
Then we have 
\begin{eqnarray}
  &&\cos(a_1)={\bf n_2} \cdot {\bf n_3},\quad 
  \cos(a_2)={\bf n_3} \cdot {\bf n_1} , \quad 
  \cos(a_3)={\bf n_1} \cdot {\bf n_2}  .
\label{A-1}
\end{eqnarray}
We make the dual unit vectors perpendicular to ${\bf n_i}$
and ${\bf n_j}$ $(i,j=1,2,3)$, $(i \ne j)$
in the following way 
\begin{eqnarray}
  &&{\bf n_1^*}=\frac{{\bf n_2}\times {\bf n_3}}
  {|{\bf n_2}\times {\bf n_3}|} , \quad
  {\bf n_2^*}=\frac{{\bf n_3}\times {\bf n_1}}
  {|{\bf n_3}\times {\bf n_1}|} , \quad
  {\bf n_3^*}=\frac{{\bf n_1}\times {\bf n_2}}
  {|{\bf n_1}\times {\bf n_2}|}  .
\label{A-2}
\end{eqnarray}
For planes perpendicular to ${\bf n_i}$ and ${\bf n_j}$, 
we denote $l_{ij}$$(i,j=1,2,3)$, $(i\ne j)$, which are given by
\begin{eqnarray}
  &&l_{23}: {\bf n_1^*}\cdot {\bf x}=0 ,\quad 
  l_{31}: {\bf n_2^*}\cdot {\bf x}=0 , \quad 
  l_{12}: {\bf n_3^*}\cdot {\bf x}=0  .
\label{A-3}
\end{eqnarray}
We denote the rotation angle of the spherical triangle around vector ${\bf n_1}$
as $A_1$ as is shown in Fig.2, then we have $\cos(A_1-\pi)={\bf n^*_2}\cdot{\bf n^*_3}$ 
from geometrical consideration.
Similarly we have 
\begin{eqnarray}
  &&\cos(A_1-\pi)={\bf n^*_2}\cdot{\bf n^*_3} , \quad
    \cos(A_2-\pi)={\bf n^*_3}\cdot{\bf n^*_1} , \quad
   \cos(A_3-\pi)={\bf n^*_1}\cdot{\bf n^*_2}  . 
\label{A-4}
\end{eqnarray} 
Using the formula 
\begin{eqnarray}
  &&({\bf a}\times {\bf b})\times ({\bf c}\times {\bf d})
    =|{\bf a},{\bf c},{\bf d}|{\bf b}-|{\bf b},{\bf c},{\bf d}|{\bf a}
    =|{\bf a},{\bf b},{\bf d}|{\bf c}-|{\bf a},{\bf b},{\bf c}|{\bf d} , 
\label{A-5}\\
  &&{\rm where} 
\nonumber\\
  &&|{\bf a},{\bf b},{\bf c}|={\bf a} \cdot ({\bf b}\times {\bf c})=
  ({\bf a}\times {\bf b})\cdot {\bf c}=({\rm determinant\ of\ }
  {\bf a},{\bf b},{\bf c})  ,
\nonumber
\end{eqnarray}
we have 
\begin{eqnarray}
  &&{\bf n_1}=\frac{{\bf n_2^*}\times {\bf n_3^*}}
  {|{\bf n_2^*}\times {\bf n_3^*}|}, \quad
  {\bf n_2}=\frac{{\bf n_3^*}\times {\bf n_1^*}}
  {|{\bf n_3^*}\times {\bf n_1^*}|}, \quad
  {\bf n_3}=\frac{{\bf n_1^*}\times {\bf n_2^*}}
  {|{\bf n_1^*}\times {\bf n_2^*}|}  .
\label{A-6}
\end{eqnarray}

\subsection{Law of cosines}
Using the theorem
\begin{eqnarray}
  &&({\bf a}\times {\bf b})\cdot({\bf c}\times {\bf d})
    =({\bf a}\cdot{\bf c})({\bf b}\cdot{\bf d})
     -({\bf a}\cdot{\bf d})({\bf b}\cdot{\bf c})  ,
\label{A-7}
\end{eqnarray}
we have 
\begin{eqnarray}
  &&\cos(\pi-A_1)={\bf n_2^*}\cdot {\bf n_3^*}
=\frac{({\bf n_3}\times {\bf n_1})}
  {|{\bf n_3}\times {\bf n_1}|} \cdot
\frac{({\bf n_1}\times {\bf n_2})}
  {|{\bf n_1}\times {\bf n_2}|}
 \nonumber\\
&&=\frac{({\bf n_1}\cdot {\bf n_3})({\bf n_1}\cdot 
 {\bf n_2})-({\bf n_2}\cdot {\bf n_3})}
 {|{\bf n_3}\times {\bf n_1}||{\bf n_1}\times {\bf n_2}|}
=\frac{\cos(a_2) \cos(a_3)-\cos(a_1)}{\sin(a_2) \sin(a_3)}  .
\label{A-8}
\end{eqnarray}
Then we have the first law of cosine in the form
\begin{eqnarray}
  && \cos(a_1)=\cos(a_2) \cos(a_3)+\cos(A_1)\sin(a_2) \sin(a_3) ,
  \nonumber\\
  && \cos(a_2)=\cos(a_3) \cos(a_1)+\cos(A_2)\sin(a_3) \sin(a_1) ,
  \nonumber\\
  && \cos(a_3)=\cos(a_1) \cos(a_2)+\cos(A_3)\sin(a_1) \sin(a_2)  .
\label{A-9}
\end{eqnarray}
While from the following relation  
\begin{eqnarray}
  &&\cos(a_1)={\bf n_2}\cdot {\bf n_3}
=\frac{({\bf n_3^*}\times {\bf n_1^*})}
  {|{\bf n_3^*}\times {\bf n_1^*}|} \cdot
\frac{({\bf n_1^*}\times {\bf n_2^*})}
  {|{\bf n_1^*}\times {\bf n_2^*}|}
=\frac{({\bf n_1^*}\cdot {\bf n_3^*})({\bf n_1^*}\cdot 
 {\bf n_2^*})-({\bf n_2^*}\cdot {\bf n_3^*})}
 {|{\bf n_3^*}\times {\bf n_1^*}||{\bf n_1^*}\times {\bf n_2^*}|}
\nonumber\\
 &&=\frac{\cos(\pi-A_2) \cos(\pi-A_3)-\cos(\pi-A_1)}{\sin(\pi-A_2) \sin(\pi-A_3)}  .
\label{A-10}
\end{eqnarray}
Then we have the second law of cosines in the following form
\begin{eqnarray}
  && -\cos(A_1)=\cos(A_2) \cos(A_3)-\cos(a_1)\sin(A_2) \sin(A_3) ,
  \nonumber\\
  && -\cos(A_2)=\cos(A_3) \cos(A_1)-\cos(a_2)\sin(A_3) \sin(A_1) ,
  \nonumber\\
  && -\cos(A_3)=\cos(A_1) \cos(A_2)-\cos(a_3)\sin(A_1) \sin(A_2) .
\label{A-11}
\end{eqnarray}

\subsection{Law of sines}
Using the relation

\begin{eqnarray}
  &&{\bf n_2^*}\times {\bf n_3^*}
=\frac{({\bf n_3}\times {\bf n_1})}
  {|{\bf n_3}\times {\bf n_1}|} \times
\frac{({\bf n_1}\times {\bf n_2})}
  {|{\bf n_1}\times {\bf n_2}|}
 \nonumber\\
&&=\frac{|{\bf n_3},{\bf n_1},{\bf n_2}|}
 {|{\bf n_3}\times {\bf n_1}||{\bf n_1}\times {\bf n_2}|} {\bf n_1}
=\frac{|{\bf n_1},{\bf n_2},{\bf n_3}|}
 {|{\bf n_3}\times {\bf n_1}||{\bf n_1}\times {\bf n_2}|} {\bf n_1}  ,
\label{A-12}
\end{eqnarray}
we have 
\begin{eqnarray}
  &&\frac{|{\bf n_2^*}\times {\bf n_3^*}|}{|{\bf n_2}\times {\bf n_3}|}
=\frac{|{\bf n_3^*}\times {\bf n_1^*}|}{|{\bf n_3}\times {\bf n_1}|}
=\frac{|{\bf n_1^*}\times {\bf n_2^*}|}{|{\bf n_1}\times {\bf n_2}|}
\nonumber\\
&&=\frac{|{\bf n_1},{\bf n_2},{\bf n_3}|}{|{\bf n_1}\times {\bf n_2}|
|{\bf n_2}\times {\bf n_3}||{\bf n_3}\times {\bf n_1}|}
=\frac{ |{\bf n_1^*},{\bf n_2^*},{\bf n_3^*}|}
{|{\bf n_1},{\bf n_2},{\bf n_3}|}  .
 \label{A-13}
\end{eqnarray}
where we use 
\begin{eqnarray}
&&|{\bf n_1^*},{\bf n_2^*},{\bf n_3^*}|=({\bf n_1^*}\times{\bf n_2^*})\cdot{\bf n_3^*}
=\frac{|{\bf n_1},{\bf n_2},{\bf n_3}|}
 {|{\bf n_2}\times {\bf n_3}||{\bf n_3}\times {\bf n_1}||{\bf n_1}\times {\bf n_2}|}
 {\bf n_3} \cdot ({\bf n_1}\times{\bf n_2})
\nonumber\\
&&=\frac{(|{\bf n_1},{\bf n_2},{\bf n_3}|)^2}
{|{\bf n_2}\times {\bf n_3}||{\bf n_3}\times {\bf n_1}||{\bf n_1}\times {\bf n_2}|}  .
 \label{A-14}
\end{eqnarray}
From Eq.(\ref{A-13}), we have the law of sines in the following form
\begin{eqnarray}
  &&\frac{\sin(\pi-A_1)}{\sin(a_1)}
=\frac{\sin(\pi-A_2)}{\sin(a_2)}
=\frac{\sin(\pi-A_3)}{\sin(a_3)} ,
\nonumber\\
&&{\rm or}
\nonumber\\
&&\frac{\sin(A_1)}{\sin(a_1)}
=\frac{\sin(A_2)}{\sin(a_2)}
=\frac{\sin(A_3)}{\sin(a_3)} .
\label{A-15}
\end{eqnarray}
We use Eq.(\ref{A-9}), Eq.(\ref{A-11}), Eq.(\ref{A-15}) 
in the body of the text.

\end{document}